\documentclass{aip-cp}

\usepackage{graphicx}

\begin{document}

%--------------------------------------------------------------------------
%                         Title and Authors
%--------------------------------------------------------------------------
\title{Dynamics of Electron Currents in Nanojunctions with Time-Varying
  Components and Interactions}

\author[aff1]{Eduardo C. Cuansing\corref{corr}}
\author[aff2]{Francis A. Bayocboc}
\author[aff3]{Christian M. Laurio}

\affil[aff1]{Institute of Mathematical Sciences and Physics, University of
  the Philippines, Los Ba\~{n}os, Laguna 4031, Philippines}
\affil[aff2]{School of Mathematics and Physics, University of Queensland,
  Queensland 4072, Australia}
\affil[aff3]{Nara Institute of Science and Technology, 8916-5 Takayama,
  Ikoma, Nara 630-0192, Japan}

\corresp[corr]{Corresponding author: eccuansing@up.edu.ph}

\maketitle

%--------------------------------------------------------------------------
%                                Abstract
%--------------------------------------------------------------------------
\begin{abstract}
  We study the dynamics of the electron current in nanodevices where there
  are time-varying components and interactions. These devices are a
  nanojunction attached to heat baths and with dynamical electron-phonon
  interactions and a nanojunction with photon beams incident and reflected
  at the channel. We use the two-time nonequilibrium Green's functions
  technique to calculate the time-dependent electron current flowing across
  the devices. We find that whenever a sudden change occurs in the device,
  the current takes time to react to the abrupt change, overshoots,
  oscillates, and eventually settles down to a steady value. With dynamical
  electron-phonon interactions, the interaction gives rise to a net
  resistance that reduces the flow of current across the device when a
  source-drain bias potential is attached. In the presence of dynamical
  electron-photon interactions, the photons drive the electrons to flow.
  The direction of flow, however, depends on the frequencies of the incident
  photons. Furthermore, the direction of electron flow in one lead is
  exactly opposite to the direction of flow in the other lead thereby
  resulting in no net change in current flowing across the device.
\end{abstract}

%--------------------------------------------------------------------------
%                        Section 1: Modeling the Nanodevices
%--------------------------------------------------------------------------
\section{MODELING THE NANODEVICES}
\label{sec:modeling}

The advent of nanometer-scale electronics has lead to compact smart devices
containing considerable computing power. As the sizes of electronic
components shrink, however, electrons in the devices are restricted to
flow within very confined dimensions. In such cases, quantum-mechanical
and nonequilibrium effects are important. In this paper, we study the
dynamics of electron currents in nanojunctions with time-varying components
and interactions. We use the two-time nonequilibrium Green's function
technique to calculate the current flowing across the
devices.\cite{haug08,stefanucci13} The nonequilibrium Green's function
technique can be used in the study of many-body quantum systems that are
partitioned into sub-parts. For example, the technique has been used in the
study of transport through quantum dots,\cite{jauho94,maciejko06,crepieux11}
a nanorelay,\cite{cuansing11} a carbon nanotube transistor,\cite{kienle10}
photon-assisted tunneling devices,\cite{henrickson02} and a nanojunction
with a time-varying gate potential.\cite{cuansing17} In this paper, we
study electron transport in a nanojunction attached to heat baths with
dynamical electron-phonon interactions and a nanojunction with an incident
photon beam striking the channel.

\begin{figure}[h!]
  \centerline{\includegraphics[width=2.6in]{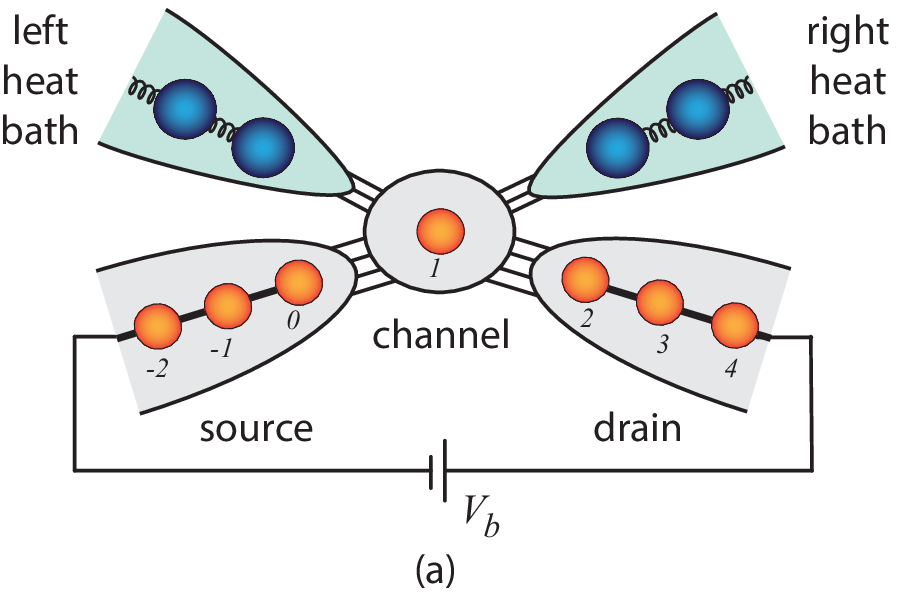}
    ~~~~~~\includegraphics[width=2.6in]{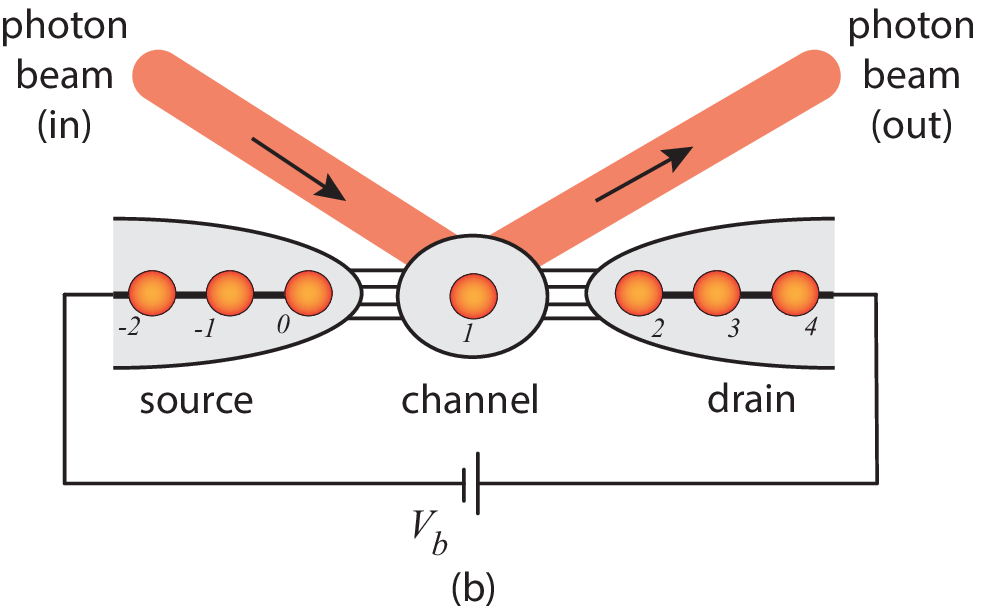}}
  \caption{An illustration of the nanojunctions with (a) attached heat baths
    and (b) an incident photon beam. Electrons flow from the source, through
    the channel, and then on to the drain. The source-drain bias potential
    is $V_b$. In (a), electrons and phonons interact within the channel. In
    (b), the incident photons interact with the electrons flowing within
    the channel.}
  \label{fig:nanojunctions}
\end{figure}

Shown in Fig.~\ref{fig:nanojunctions}(a) is an illustration of a nanojunction
with left and right heat baths attached to the channel. The heat baths have
temperatures $T_{HB}^{\rm L} > T_{HB}^{\rm R}$ and therefore phonons flow from
the left to the right heat bath. The direction of electron flow, on the other
hand, depends on the sign of the source-drain bias potential $V_b$. We call
the source as the left lead and the drain as the right lead. The left and
right leads have chemical potentials $\mu_{\rm L}$ and $\mu_{\rm R}$ and
temperatures $T_e^{\rm L}$ and $T_e^{\rm R}$. We use the tight-binding
approximation to model the devices. The Hamiltonian for the left and right
leads are $H_e^{\rm L} = \sum \varepsilon_k^{\rm L}\,a_k^{\dagger} a_k +
\sum v_{kj}^{\rm L}\,\left( a_k^{\dagger} a_j + a_j^{\dagger} a_k\right)$ and
$H_e^{\rm R} = \sum \varepsilon_k^{\rm R}\,b_k^{\dagger} b_k + \sum v_{kj}^{\rm R}
\,\left( b_k^{\dagger} b_j + b_j^{\dagger} b_k\right)$, where $a_k$ and
$a_k^{\dagger}$ are the electron annihilation and creation operators at site
$k$ in the left lead, $b_k$ and $b_k^{\dagger}$ are the annihilation and
creation operators at site $k$ in the right lead, the $\varepsilon_k^{\rm L}$
and $\varepsilon_k^{\rm R}$ are the on-site energies at site $k$, the
$v_{kj}^{\rm L}$ and $v_{kj}^{\rm R}$ are the hopping parameters for
nearest-neighbor sites $k$ and $j$, and the sums are over all sites in the
leads. The leads are connected to the channel with the leads-channel
coupling Hamiltonian $H_e^{\rm LC} = v_{01}^{\rm LC}\left(a_0^{\dagger} c_1 +
c_1^{\dagger} a_0\right)$ and $H_e^{\rm RC} = v_{21}^{\rm RC}\left(c_1^{\dagger}
b_2 + b_2^{\dagger} c_1\right)$, where $v_{01}^{\rm LC} = v_{10}^{\rm CL}$ and
$v_{21}^{\rm RC} = v_{12}^{\rm CR}$ are the symmetric leads-channel coupling
parameters. The Hamiltonian for the phonons in the left and right heat baths
are $H_p^{\rm L} = \sum \hbar k_k^{\rm L}\,q_k^{\dagger} q_k + \sum \hbar
\kappa_{jk}^{\rm L}\,q_j^{\dagger} q_k$ and $H_p^{\rm R} = \sum \hbar k_k^{\rm R}\,
r_k^{\dagger} r_k + \sum \hbar \kappa_{jk}^{\rm R}\,r_j^{\dagger} r_k$, where
$\hbar k_k^{\rm L}$ and $\hbar k_k^{\rm R}$ are the on-site energies of the
phonons at site $k$ in the heat baths, $\hbar\kappa_{jk}^{\rm L}$ and
$\hbar\kappa_{jk}^{\rm R}$ are the phonon hopping parameters between
nearest-neighbor sites $j$ and $k$, $q_k^{\dagger}$ and $q_k$ are the phonon
creation and annihilation operators at site $k$ in the left heat bath, and
$r_k^{\dagger}$ and $r_k$ are the phonon creation and annihilation operators
at site $k$ in the right heat bath. The heat baths are coupled to the
channel via $H_p^{\rm LC} = \hbar\kappa_{01}^{\rm LC}\left( q_0^{\dagger} p_1
+ p_1^{\dagger} q_0\right)$ and $H_p^{\rm RC} = \hbar\kappa_{21}^{\rm RC}\left(
r_2^{\dagger} p_1 + p_1^{\dagger} r_2\right)$, where $p_1^{\dagger}$ and $p_1$
are the phonon creation and annihilation operators in the channel. The
Hamiltonian in the channel contains an electron part $H_e^{\rm C} =
\varepsilon_1^{\rm C}\,c_1^{\dagger} c_1$, a phonon part $H_p^{\rm C} =
\hbar k_1^{\rm C}\,p_1^{\dagger} p_1$, and the time-dependent electron-phonon
interaction term $H_{ep}^{\rm C}(t) = M_{ep}(t)\left(p_1^{\dagger} + p_1\right)
c_1^{\dagger} c_1$, where $M_{ep}(t)$ is the electron-phonon interaction
coupling parameter.\cite{ziman72,madelung78,lu07} The total Hamiltonian of
the device is $H = H_e + H_p + H_{ep}(t)$, where the time-independent electron
part is $H_e = H_e^{\rm L} + H_e^{\rm R} + H_e^{\rm C} + H_e^{\rm LC} +
H_e^{\rm RC}$, the time-independent phonon part is $H_p = H_p^{\rm L} +
H_p^{\rm R} + H_p^{\rm C} + H_p^{\rm LC} + H_p^{\rm RC}$, and the time-dependent
electron-phonon interaction term is in $H_{ep}(t) = H_{ep}^{\rm C}(t)$.

We also study a nanojunction where a photon beam is incident on the channel,
as illustrated in Fig.~\ref{fig:nanojunctions}(b). The time-independent
electron part of the Hamiltonian, $H_e$, is the same as in the previous device.
The time-independent photon part of the Hamiltonian includes
$H_{\gamma} = H_{\gamma}^{\rm L} + H_{\gamma}^{\rm R} + H_{\gamma}^{\rm C} +
H_{\gamma}^{\rm LC} + H_{\gamma}^{\rm RC}$. The Hamiltonian in the incident and
reflected photon beams are $H_{\gamma}^{\rm L} = \sum \chi_k^{\rm L}\,
\alpha_k^{\dagger} \alpha_k + \sum u_{jk}^{\rm L}\,\alpha_j^{\dagger} \alpha_k$
and $H_{\gamma}^{\rm R} = \sum \chi_k^{\rm R}\,\beta_k^{\dagger} \beta_k +
\sum u_{jk}^{\rm R}\,\beta_j^{\dagger} \beta_k$, where $\chi_k^{\rm L}$ and
$\chi_k^{\rm R}$ are the photon on-site energy at site $k$ in the beams,
$u_{jk}^{\rm L}$ and $u_{jk}^{\rm R}$ are the photon hopping parameters for
nearest neighbors $j$ and $k$, $\alpha_k^{\dagger}$ and $\alpha_k$ are the
photon creation and annihilation operators in the left incident beam,
$\beta_k^{\dagger}$ and $\beta_k$ are the photon creation and annihilation
operators in the right reflected beam, and the sums are over all sites in
the photon beams. The beam-channel coupling parts of the Hamiltonian
are $H_{\gamma}^{\rm LC} = u_{01}^{\rm LC}\left( \alpha_0^{\dagger} \gamma_1
+ \gamma_1^{\dagger} \alpha_0\right)$ and $H_{\gamma}^{\rm RC} = u_{21}^{\rm RC}
\left( \beta_2^{\dagger} \gamma_1 + \gamma_1^{\dagger} \beta_2\right)$, where
$\gamma_1^{\dagger}$ and $\gamma_1$ are the photon creation and annihilation
operators at the site in the channel. The channel part of the Hamiltonian
contains the time-independent electron part $H_e^{\rm C} = \varepsilon_1^{\rm C}
\,c_1^{\dagger} c_1$, the time-independent photon part $H_{\gamma}^{\rm C} =
\chi_1^{\rm C}\,\gamma_1^{\dagger} \gamma_1$, and the part containing the
time-dependent electron-photon interaction $H_{e\gamma}^{\rm C}(t) =
M_{e\gamma}(t)\left(\gamma_1 e^{-i\omega t} + \gamma_1^{\dagger}
e^{i\omega t}\right) c_1^{\dagger} c_1$, where $M_{e\gamma}(t)$ is the
electron-photon interaction strength and $\omega$ is the frequency of the
photon.\cite{henrickson02,madelung78,galperin12} The total Hamiltonian for
the device is $H = H_e + H_{\gamma} + H_{e\gamma}(t)$ where $H_e$ is the
time-independent electron part, $H_{\gamma}$ is the time-independent photon
part, and $H_{e\gamma}(t) = H_{e\gamma}^{\rm C}(t)$ is the time-dependent
electron-photon interaction term in the channel.

The electron current flowing out of the left lead can be determined from the
time rate of change of the number of electrons in the lead. Let $N^{\rm L} =
\sum a_k^{\dagger} a_k$ be the number operator in the left lead. Then, the
current out of the left lead is
\begin{equation}
  I^{\rm L}(t) = \left\langle - q \frac{d N^{\rm L}}{d t}\right\rangle
  = -\frac{i q}{\hbar} \left\langle \left[ H, N^{\rm L}\right]\right\rangle
  = -\frac{i q}{\hbar} v_{01}^{\rm LC} \left(-\langle a_0^{\dagger}
  c_1\rangle + \langle c_1^{\dagger} a_0\rangle\right) = 2 q\,{\rm Re}\!
  \left[ v_{01}^{\rm LC} G_{10}^{{\rm CL},<}(t,t)\right],
  \label{eq:Ileft}
\end{equation}
where the Heisenberg equation of motion is used in the second equality, the
${\rm Re}[\,]$ indicates the real part, and the lesser nonequilibrium Green's
function is $G_{10}^{{\rm CL},<}(t_1,t_2) = \frac{i}{\hbar}\left\langle
a_0^{\dagger}(t_2)\,c_1(t_1)\right\rangle$. Similarly, the electron current
flowing out of the right lead is
\begin{equation}
  I^{\rm R}(t) = \left\langle q \frac{d N^{\rm R}}{d t}\right\rangle
  = -2 q\,{\rm Re}\!\left[ v_{21}^{\rm RC} G_{12}^{{\rm CR},<}(t,t)\right],
  \label{eq:Iright}
\end{equation}
where the lesser nonequilibrium Green's function $G_{12}^{{\rm CR},<}(t_1,t_2)
= \frac{i}{\hbar}\left\langle b_2^{\dagger}(t_2)\,c_1(t_1)\right\rangle$. The
results are the same for the two Hamiltonians we defined for the two types
of nanojunctions we are studying. Note that in both Eqs.~(\ref{eq:Ileft})
and (\ref{eq:Iright}) a positive current means it is flowing into the right
lead while a negative current means it is flowing into the left lead.

The lesser nonequilibrium Green's functions needed to determine the left
and right currents can be calculated following the Keldysh-Schwinger
formalism where time is on a Keldysh contour in a complex contour-time
plane.\cite{haug08,stefanucci13} We start with the contour-ordered Green's
function $G_{10}^{\rm CL}(\tau_1,\tau_2) = -\frac{i}{\hbar}\left\langle
{\rm T}_c\,c_1(\tau_1)\,a_0^{\dagger}(\tau_2)\right\rangle$, where $\tau_1$
and $\tau_2$ are contour-time variables and ${\rm T}_c$ is the contour-ordering
operator. In the Interaction Picture, the contour-ordered Green's function
becomes $G_{10}^{\rm CL}(\tau_1,\tau_2) = -\frac{i}{\hbar}\left\langle
{\rm T}_c\,e^{-\frac{i}{\hbar} \int_c H_t(\tau')\,d\tau'} c_1(\tau_1)\,
a_0^{\dagger}(\tau_2)\right\rangle_0$, where the angled brackets
$\langle\,\rangle_0$ indicate ensemble averages with respect to the steady
state and the Hamiltonian is separated into $H = H_0 + H_t$, where $H_0$ is
the stationary part and $H_t$ is the time-dependent part. In the nanojunction
attached to heat baths, for example, we have $H_t = H_{ep}^{\rm C}(t)$ while
in the nanojunction with an incident photon beam $H_t = H_{e\gamma}^{\rm C}(t)$.

\begin{figure}[h!]
  \centerline{\includegraphics[width=4.5in]{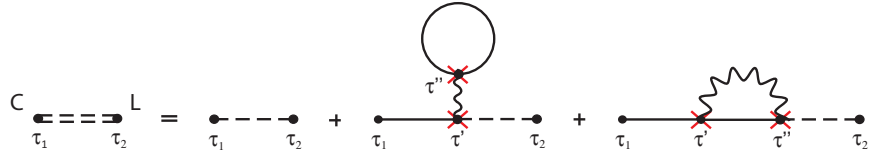}}
  \caption{Connected diagrams for the perturbation expansion of
    $G_{10}^{\rm CL}(\tau_1,\tau_2)$, up to the second-order terms. The wavy
    lines are the phonon steady-state Green's funcition and each $3$-vertex
    represents an electron-phonon interaction coupling.}
  \label{fig:diagrams}
\end{figure}

Consider the nanojunction that is attached to heat baths. The perturbation
expansion of the contour-ordered Green's function
$G_{10}^{\rm CL}(\tau_1,\tau_2)$ using $H_{ep}^{\rm C}(t)$ as the perturbing
Hamiltonian results in the diagrammatic expansion shown in
Fig.~\ref{fig:diagrams}, up to the second-order term. The expansion translates
to
\begin{eqnarray}
  G_{10}^{\rm CL}(\tau_1,\tau_2) & = & G_{10,0}^{\rm CL}(\tau_1,\tau_2)
  - \int_c d\tau' \int_c d\tau''\,G_{11,0}^{\rm CC}(\tau_1,\tau')\,
  M_{ep}(\tau')\,D_{11,0}^{\rm CC}(\tau',\tau'')\,G_{11,0}^{\rm CC}(\tau'',\tau'')
  \,M_{ep}(\tau'')\,G_{10,0}^{\rm CL}(\tau',\tau_2)\nonumber\\
  & ~ & + \int_c d\tau' \int_c d\tau''\,G_{11,0}^{\rm CC}(\tau_1,\tau')\,
  M_{ep}(\tau')\,D_{11,0}^{\rm CC}(\tau',\tau'')\,G_{11,0}^{\rm CC}(\tau',\tau'')
  \,M_{ep}(\tau'')\,G_{10,0}^{\rm CL}(\tau'',\tau_2),
  \label{eq:GCL}
\end{eqnarray}
where the extra $0$ subscript indicates a steady-state Green's function and
the phonon steady-state Green's function in the channel
$D_{11,0}^{\rm CC}(\tau_1,\tau_2) = -\frac{i}{\hbar} \left\langle {\rm T}_c
p_1(\tau_1) p_1^{\dagger}(\tau_2)\right\rangle_0$. The diagram containing the
fermionic loop vanishes upon integration along the Keldysh contour. We then
perform analytic continuation and use Langreth's
theorem\cite{haug08,stefanucci13} to Eq.~(\ref{eq:GCL}) to determine the
associated nonequilibrium Green's functions in real time. The lesser
nonequilibrium Green's function becomes
\begin{eqnarray}
  G_{10}^{{\rm CL},<}(t_1,t_2) & = & G_{10,0}^{{\rm CL},<}(t_1,t_2) + \int_0^t
  dt' \int_0^t dt''\,G_{11,0}^{{\rm CC},r}(t_1,t')\,M_{ep}(t')\,
  F_{ep}^r(t',t'')\,M_{ep}(t'')\,G_{10,0}^{{\rm CL},<}(t'',t_2)\nonumber\\
  & ~ & + \int_0^t dt' \int_0^t dt''\,G_{11,0}^{{\rm CC},r}(t_1,t')\,
  M_{ep}(t')\,F_{ep}^<(t',t'')\,M_{ep}(t'')\,G_{10,0}^{{\rm CL},a}(t'',t_2)\\
  & ~ & + \int_0^t dt' \int_0^t dt''\,G_{11,0}^{{\rm CC},<}(t_1,t')\,
  M_{ep}(t')\,F_{ep}^a(t',t'')\,M_{ep}(t'')\,G_{10,0}^{{\rm CL},a}(t'',t_2),
  \nonumber
  \label{eq:GCLless}
\end{eqnarray}
where
\begin{eqnarray}
  F_{ep}^r(t',t'') & = & D_{11,0}^{{\rm CC},<}(t',t'')\,G_{11,0}^{{\rm CC},r}(t',t'')
  + D_{11,0}^{{\rm CC},r}(t',t'')\,G_{11,0}^{{\rm CC},<}(t',t'') +
  D_{11,0}^{{\rm CC},r}(t',t'')\,G_{11,0}^{{\rm CC},r}(t',t''),\nonumber\\
  F_{ep}^<(t',t'') & = & D_{11,0}^{{\rm CC},<}(t',t'')\,
  G_{11,0}^{{\rm CC},<}(t',t''),\\
  F_{ep}^a(t',t'') & = & \left[ F_{ep}^r(t',t'')\right]^{\ast}.\nonumber
  \label{eq:Fepr}
\end{eqnarray}
  
For the nanojunction with an incident photon beam, we can perform a
perturbation expansion of $G_{10}^{\rm CL}(\tau_1,\tau_2)$ with
$H_{e\gamma}^{\rm C}(t)$ as the perturbing Hamiltonian. The diagrammatic
expansion results in the same diagrams shown in Fig.~\ref{fig:diagrams}, up
to the second-order terms, except for the reinterpretation of all wavy
lines as photon steady-state Green's functions. There are, however, two
versions of each diagram because of the $e^{-i\omega t}$ and $e^{i\omega t}$
coefficients in the electron-photon interaction terms. One diagram is for
a forward-moving photon while the other is for a backward-moving photon.
Diagrams with fermionic loops do not contribute to the expansion. For this
nanojunction, the lesser nonequilibrium Green's function is
\begin{eqnarray}
  G_{10}^{{\rm CL},<}(t_1,t_2) & = & G_{10,0}^{{\rm CL},<}(t_1,t_2)
  + \int_0^t dt' \int_0^t dt''\,G_{11,0}^{{\rm CC},r}(t_1,t')\,M_{e\gamma}(t')
  \left[ F_1^r(t',t'') + F_2^r(t',t'')\right] M_{e\gamma}(t'')\,
  G_{10,0}^{{\rm CL},<}(t'',t_2)\nonumber\\
  & ~ & + \int_0^t dt' \int_0^t dt''\,G_{11,0}^{{\rm CC},r}(t_1,t')\,
  M_{e\gamma}(t') \left[ F_1^<(t',t'') + F_2^<(t',t'')\right] M_{e\gamma}(t'')\,
  G_{10,0}^{{\rm CL},a}(t'',t_2)\\
  & ~ & + \int_0^t dt' \int_0^t dt''\,G_{11,0}^{{\rm CC},<}(t_1,t')\,
  M_{e\gamma}(t') \left[ F_1^a(t',t'') + F_2^a(t',t'')\right] M_{e\gamma}(t'')\,
  G_{10,0}^{{\rm CL},a}(t'',t_2),\nonumber
  \label{eq:GCLlessphoton}
\end{eqnarray}
where
\begin{eqnarray}
  F_1^r(t',t'') & = & G_{11,0}^{{\rm CC},r}(t',t'')\,
  \tilde{D}_{11,0}^{{\rm CC},<}(t',t'') + G_{11,0}^{{\rm CC},<}(t',t'')\,
  \tilde{D}_{11,0}^{{\rm CC},r}(t',t'') + G_{11,0}^{{\rm CC},r}(t',t'')\,
  \tilde{D}_{11,0}^{{\rm CC},r}(t',t''),\\
  F_2^r(t',t'') & = & G_{11,0}^{{\rm CC},<}(t',t'')\,
  \tilde{D}_{11,0}^{{\rm CC},a}(t'',t') + G_{11,0}^{{\rm CC},r}(t',t'')\,
  \tilde{D}_{11,0}^{{\rm CC},<}(t'',t'),
\end{eqnarray}
$F_1^a(t',t'') = \left[ F_1^r(t',t'')\right]^{\dagger}$,
$F_2^a(t',t'') = \left[ F_2^r(t',t'')\right]^{\dagger}$,
$F_1^<(t',t'') = G_{11,0}^{{\rm CC},<}(t',t'')\,
\tilde{D}_{11,0}^{{\rm CC},<}(t',t'')$, and $F_2^<(t',t'') =
G_{11,0}^{{\rm CC},<}(t',t'')\,\tilde{D}_{11,0}^{{\rm CC},>}(t'',t')$, where
$\tilde{D}_{11,0}^{{\rm CC},>}(t',t'') = \left[
\tilde{D}_{11,0}^{{\rm CC},<}(t',t'')\right]^{\ast}$, the modified photon
steady-state Green's function $\tilde{D}_{11,0}^{{\rm CC},\alpha}(t',t'') =
e^{-i\omega t'}D_{11,0}^{{\rm CC},\alpha}(t',t'')\,e^{i\omega t''}$ and
$D_{11,0}^{{\rm CC},\alpha}(t',t'')$ is the steady-state photon Green's
function with $\alpha = r, a, <, >$. 

The electron steady-state Green's functions can be built from the adiabatic
switch-on of the coupling between the leads and the channel. Since
time-translation invariance is satisfied in the steady-state regime, the
Fourier transforms of the steady-state Green's functions are well-defined.
For the electron ${\rm CL}$ steady-state Green's functions we get, in
Fourier space,
\begin{eqnarray}
  G_{10,0}^{{\rm CL},<}(E) & = & G_{11,0}^{{\rm CC},r}(E)\,v_{10}^{\rm CL}\,
  g_{00}^{{\rm L},<}(E) + G_{11,0}^{{\rm CC},<}(E)\,v_{10}^{\rm CL}\,
  g_{00}^{{\rm L},a}(E),\nonumber\\
  G_{10,0}^{{\rm CL},r}(E) & = & G_{11,0}^{{\rm CC},r}(E)\,v_{10}^{\rm CL}\,
  g_{00}^{{\rm L},r}(E),\\
  G_{10,0}^{{\rm CL},a}(E) & = & \left[ G_{10,0}^{{\rm CL},r}(E)\right]^{\ast},
  \nonumber
  \label{eq:GCLsteady}
\end{eqnarray}
where the $g_{00}^{{\rm L},\alpha}(E)$, $\alpha=r,a,<$, are the equilibrium
Green's functions in the left lead. Furthermore, the electron ${\rm CC}$
steady-state Green's functions are
\begin{eqnarray}
  G_{11,0}^{{\rm CC},<}(E) & = & G_{11,0}^{{\rm CC},r}(E)\,
  \Sigma_{e,11}^{{\rm C},<}(E)\,G_{11,0}^{{\rm CC},a}(E),\nonumber\\
  G_{11,0}^{{\rm CC},r}(E) & = & \left[ (E + i\eta)-\varepsilon_1^{\rm C} -
    \Sigma_{e,11}^{{\rm C},r}(E)\right]^{-1},\\
  G_{11,0}^{{\rm CC},a}(E) & = & \left[ G_{11,0}^{{\rm CC},r}(E)\right]^{\ast},
  \nonumber
  \label{eq:GCCsteady}
\end{eqnarray}
where the electron self-energy $\Sigma_{e,11}^{{\rm C},\alpha}(E) =
v_{10}^{\rm CL}\,g_{00}^{{\rm L},\alpha}(E)\,v_{01}^{\rm LC} + v_{12}^{\rm CR}\,
g_{22}^{{\rm R},\alpha}(E)\,v_{21}^{\rm RC}$ and the $g_{22}^{{\rm R},\alpha}(E)$
are the equilibrium Green's functions in the right lead. The phonon
steady-state Green's functions have the same form as the electron
steady-state Green's functions except for the substitution of
$E\rightarrow\hbar\omega$, $\varepsilon_1^{\rm C}\rightarrow\hbar k_1^{\rm C}$,
$v_{10}^{\rm CL}\rightarrow\hbar\kappa_{10}^{\rm CL}$, $v_{12}^{\rm CR}\rightarrow
\hbar\kappa_{12}^{\rm CR}$, and the electron self-energy
$\Sigma_{e,11}^{{\rm C},\alpha}(E)$ to the phonon self-energy
$\Sigma_{p,11}^{{\rm C},\alpha}(E) = \hbar\kappa_{10}^{\rm CL}\,
d_{00}^{{\rm L},\alpha}(E)\,\hbar\kappa_{01}^{\rm LC} + \hbar\kappa_{12}^{\rm CR}\,
d_{22}^{{\rm R},\alpha}(E)\,\hbar\kappa_{21}^{\rm RC}$, where the
$d_{00}^{{\rm L},\alpha}(E)$ and $d_{22}^{{\rm R},\alpha}(E)$ are the phonon
equilibrium Green's functions in the left and right heat baths, respectively.
The photon steady-state Green's functions are also of the same form as the
electron steady-state Green's functions except for the substitution of
$\varepsilon_1^{\rm C}\rightarrow\chi_1^{\rm C}$,
$v_{10}^{\rm CL}\rightarrow u_{10}^{\rm CL}$, $v_{12}^{\rm CR}\rightarrow
u_{12}^{\rm CR}$, and the electron self-energy $\Sigma_{e,11}^{{\rm C},\alpha}(E)$
to the photon self-energy $\Sigma_{\gamma,11}^{{\rm C},\alpha}(E) = u_{10}^{\rm CL}
\,p_{00}^{{\rm L},\alpha}(E)\,u_{01}^{\rm LC} + u_{12}^{\rm CR}\,
p_{22}^{{\rm R},\alpha}(E)\, u_{21}^{\rm RC}$, where the $p_{00}^{{\rm L},\alpha}(E)$
and $p_{22}^{{\rm R},\alpha}(E)$ are the photon equilibrium Green's functions
in the left and right photon beams.

The electron equilibrium Green's functions can be determined from the
equation of motion of the electrons in the leads.\cite{haug08,stefanucci13}
For the left lead, we get
\begin{eqnarray}
  g_{00}^{{\rm L},r}(E) & = & 2\,\frac{(E+i\eta)-\varepsilon_0^{\rm L}}{v^2}
  \pm 2 i\,\frac{\sqrt{v^2-(\varepsilon_0^{\rm L}-E)^2}}{v^2},\nonumber\\
  g_{00}^{{\rm L},a}(E) & = & \left[ g_{00}^{{\rm L},r}(E)\right]^{\ast},\\
  g_{00}^{{\rm L},<}(E) & = & - f^{\rm L}(E) \left(g_{00}^{{\rm L},r}(E)-
  g_{00}^{{\rm L},a}(E)\right),\nonumber
  \label{eq:gL}
\end{eqnarray}
where $f^{\rm L}(E)=\left[e^{(E-\mu_{\rm L})/k_B T_e^{\rm L}}+1\right]^{-1}$ is the
Fermi-Dirac distribution function of the left lead. The equilibrium Green's
functions for the right lead are of the same form as those for the left
lead except for the replacement of all left-lead parameters into right-lead
parameters. The phonon equilibrium Green's functions are of the same form
as the electron equilibrium Green's functions except for the substitution
of all electron parameters into phonon parameters, the use of the
Bose-Einstein distribution function, and taking into consideration that
phonon operators follow the commutation rule for bosons. The photon
equilibrium Green's functions are of the same form as the phonon equilibrium
Green's functions except for the substition of all phonon parameters to
photon parameters.

%--------------------------------------------------------------------------
%                      Section 2: Numerical Results
%--------------------------------------------------------------------------
\section{NUMERICAL RESULTS}
\label{sec:results}

To calculate the time-dependent left and right currents we need to determine
the lesser nonequilibrium Green's functions $G_{10}^{{\rm CL},<}(t_1,t_2)$
and $G_{12}^{{\rm CR},<}(t_1,t_2)$. These nonequilibrium Green's functions
are expressed as integrals of steady-state Green's functions, as shown in
Eqs.~(\ref{eq:GCLless}) and (\ref{eq:GCLlessphoton}). We calculate these
integrals numerically using standard integration techniques.\cite{press07}

In our calculations, the electron parameters we use are $\varepsilon^{\rm L}
= \varepsilon^{\rm R} = \varepsilon^{\rm C} = 0$ and $v^{\rm L} = v^{\rm R}
= v^{\rm C} = v^{\rm LC} = v^{\rm RC} = 2~{\rm eV}$. The left lead parameters
are $\mu_{\rm L} = E_f$ and $T_e^{\rm L} = 300~{\rm K}$ while the right lead
parameters are $\mu_{\rm R} = E_f-V_b$ and $T_e^{\rm R} = 300~{\rm K}$, where
the Fermi energy $E_f = 0$ and $V_b$ is the source-drain bias potential.

\begin{figure}[h!]
  \centerline{\includegraphics[width=2.8in,clip]{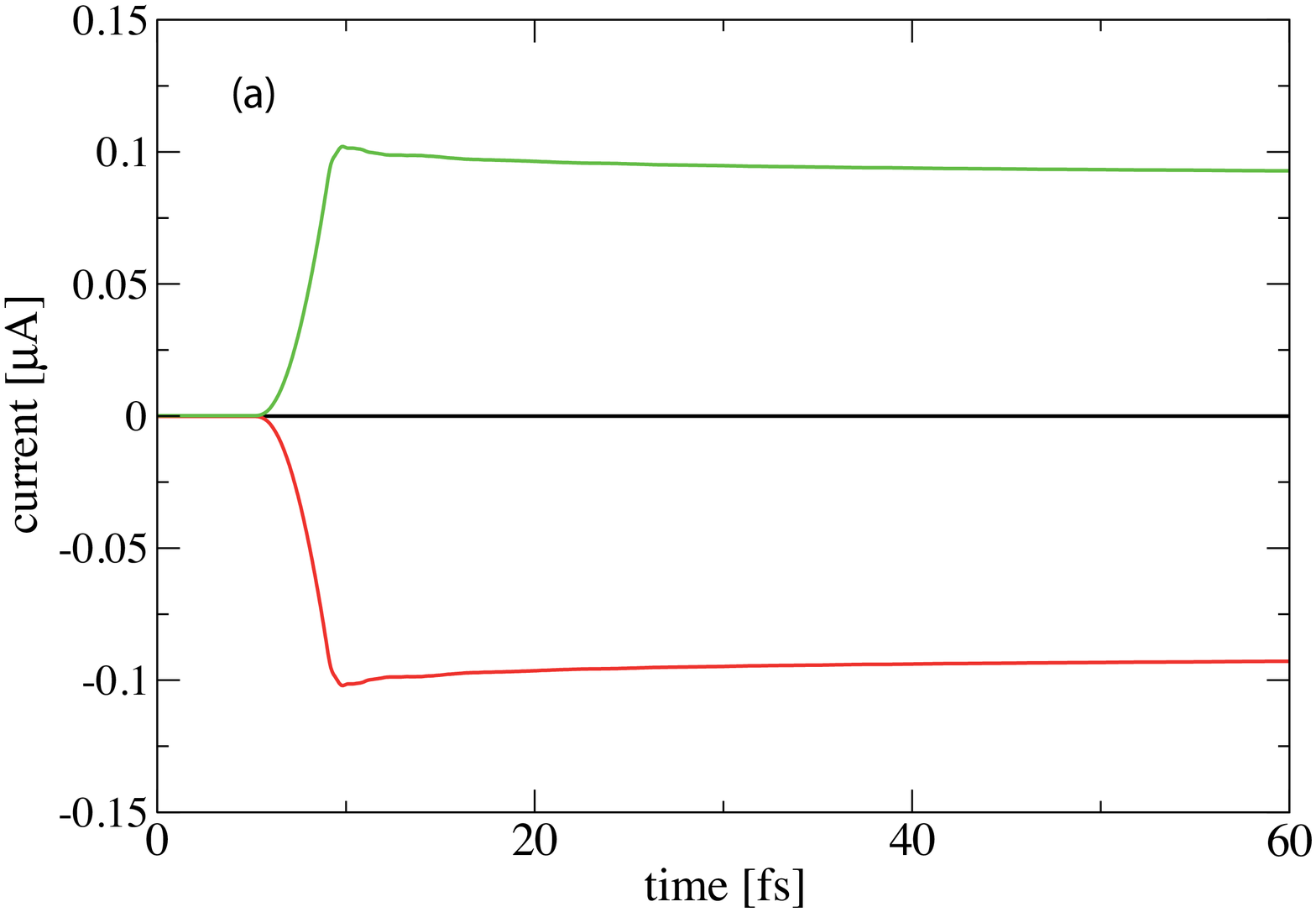}~~~~~
    \includegraphics[width=2.8in,clip]{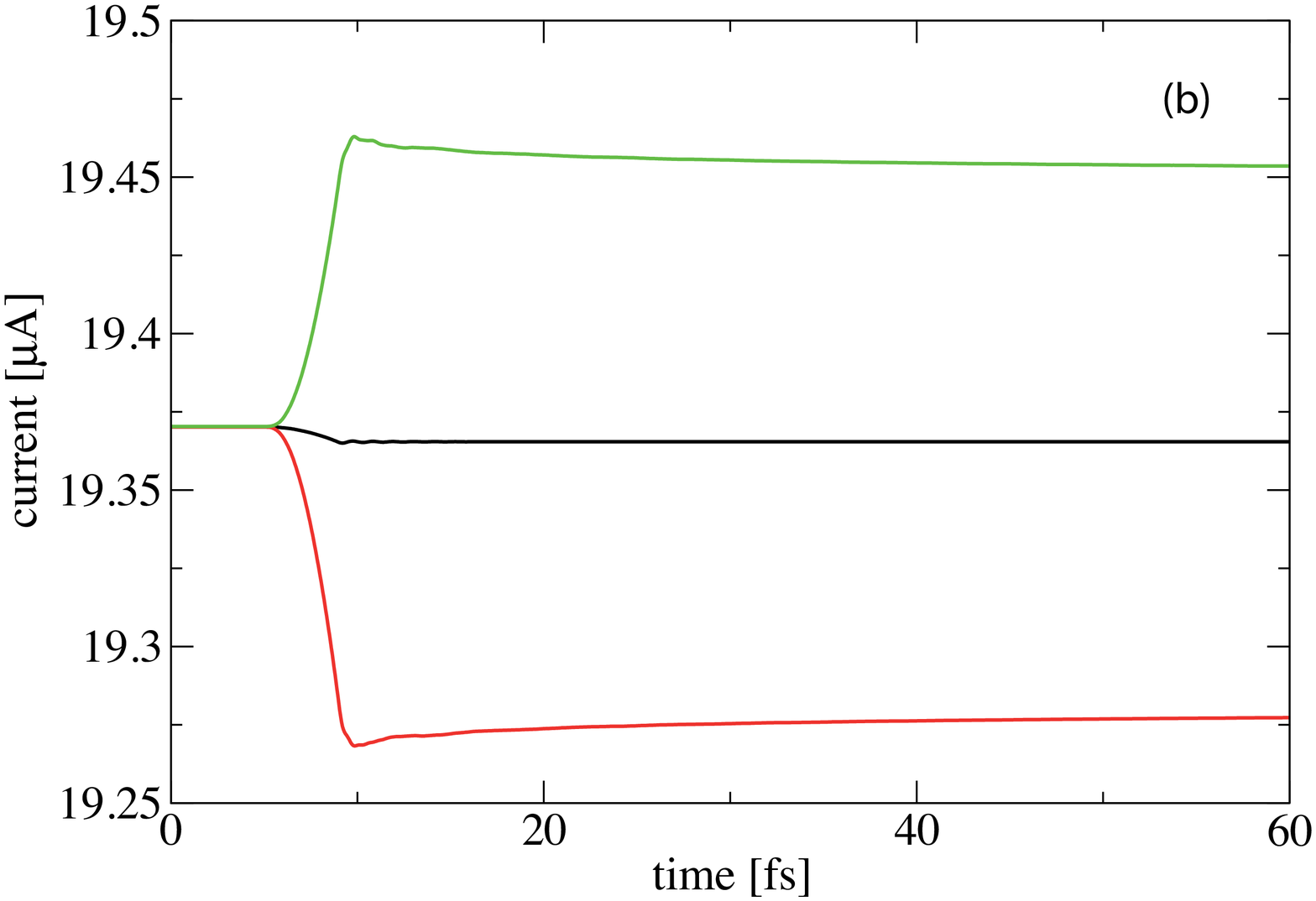}}
  \caption{The left current (dark line, red online), right current (light
    line, green online), and average current (black line) for the
    nanojunction with electron-phonon interactions in the channel is
    switched on abruptly at $4~{\rm fs}$. The source-drain bias potential
    in (a) is $V_b = 0$ while in (b) it is $V_b = 0.5~{\rm eV}$.}
  \label{fig:electronphonon}
\end{figure}

Shown in Fig.~\ref{fig:electronphonon} are the left and right currents
in the nanojunction that is attached to heat baths. The phonon parameters
we use are $\hbar k^{\rm L} = \hbar k^{\rm R} = \hbar k^{\rm C} = 0.2~{\rm eV}$
and $\hbar\kappa^{\rm L} = \hbar\kappa^{\rm R} = \hbar\kappa^{\rm C} = \hbar
\kappa^{\rm LC} = \hbar\kappa^{\rm RC} = 0.5~{\rm eV}$. The left and right
heat bath temperatures are $T_{HB}^{\rm L} = 473~{\rm K}$ and $T_{HB}^{\rm R}
= 273~{\rm K}$ and the electron-phonon coupling strength is $M_{ep} =
0.5~{\rm eV}$. Electron-phonon interactions are switched on abruptly, i.e.,
in the form of a Heaviside step function, at time $t = 4~{\rm fs}$. Because
$T_{HB}^{\rm L} > T_{HB}^{\rm R}$, phonons flow from the left heat bath to the
right bath. At the channel, the interactions between the electrons and phonons
affect the electron current across the device. Shown in
Fig.~\ref{fig:electronphonon}(a) are the left and right currents when there
is no source-drain bias potential. Since $V_b = 0$, there initially is no
current flowing. Then when the electron-phonon interaction is suddenly
switched on, electrons start to flow in such a way that the left current is
directed towards the left lead and the right current is directed towards the
right lead, i.e., the direction of current flow is outwards from the channel,
in both directions. The average current $I_{ave} = \frac{1}{2} \left(
I^{\rm L}(t)+I^{\rm R}(t)\right)$ remains zero since the form and sign of the
left current are exactly the opposite of the right current. Notice also that
the currents do not instantly flow when the electron-phonon interaction is
abruptly switched on but instead they take time to react, overshoots, and
then settle down to steady values. When the nanojunction is attached
to a source-drain bias potential $V_b = 0.5~{\rm eV}$, we get the currents
shown in Fig.~\ref{fig:electronphonon}(b). The left and right currents
appear to have the same form. However, the decrease in the left current
is more than the increase in the right current thereby resulting in an
average current that is less than its original value. The electron-phonon
interaction, therefore, gives rise to a resistance that reduces the net
current flowing across the device.

\begin{figure}[h!]
  \centerline{\includegraphics[width=3in,clip]{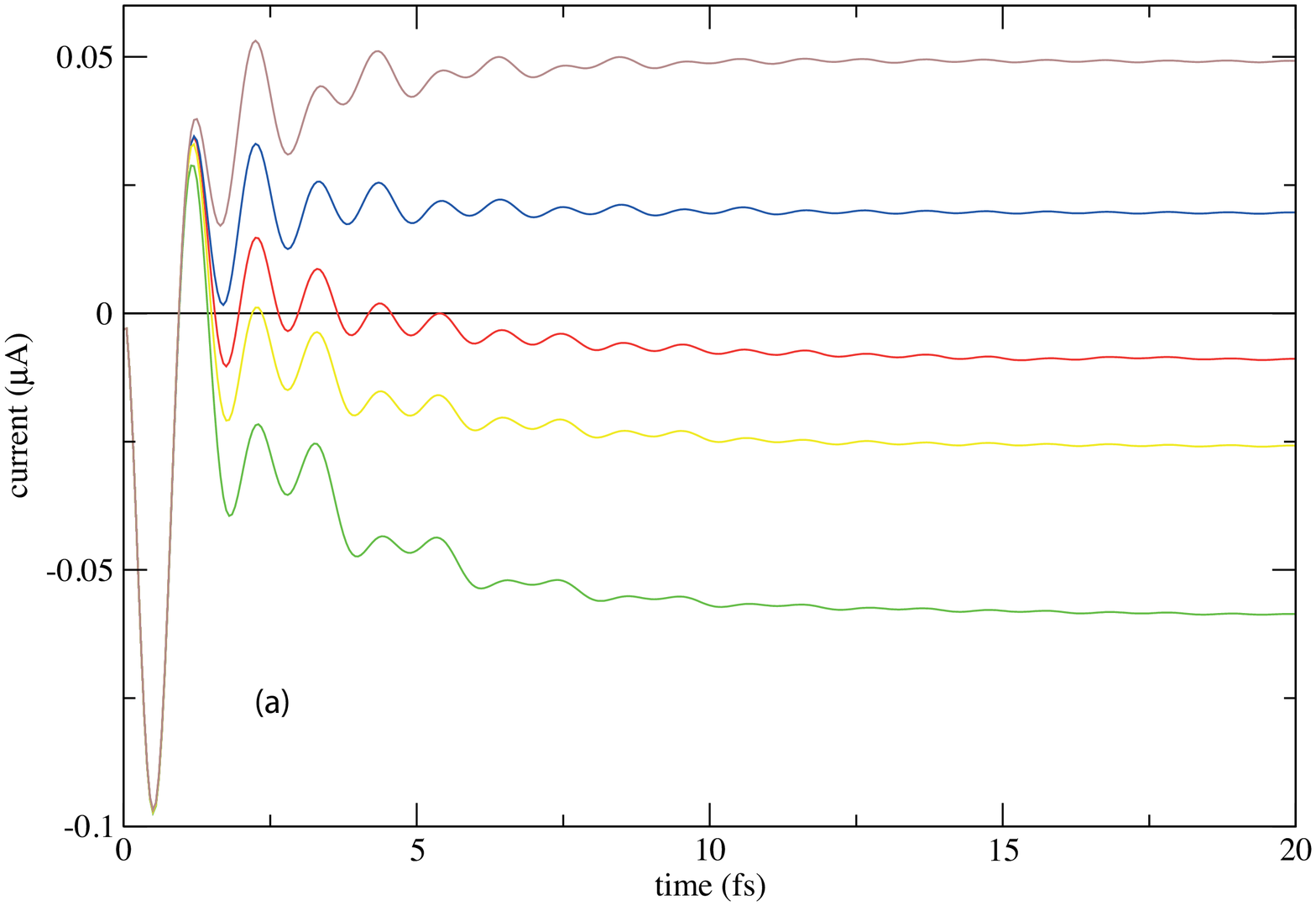}
  ~~~~\includegraphics[width=3in,clip]{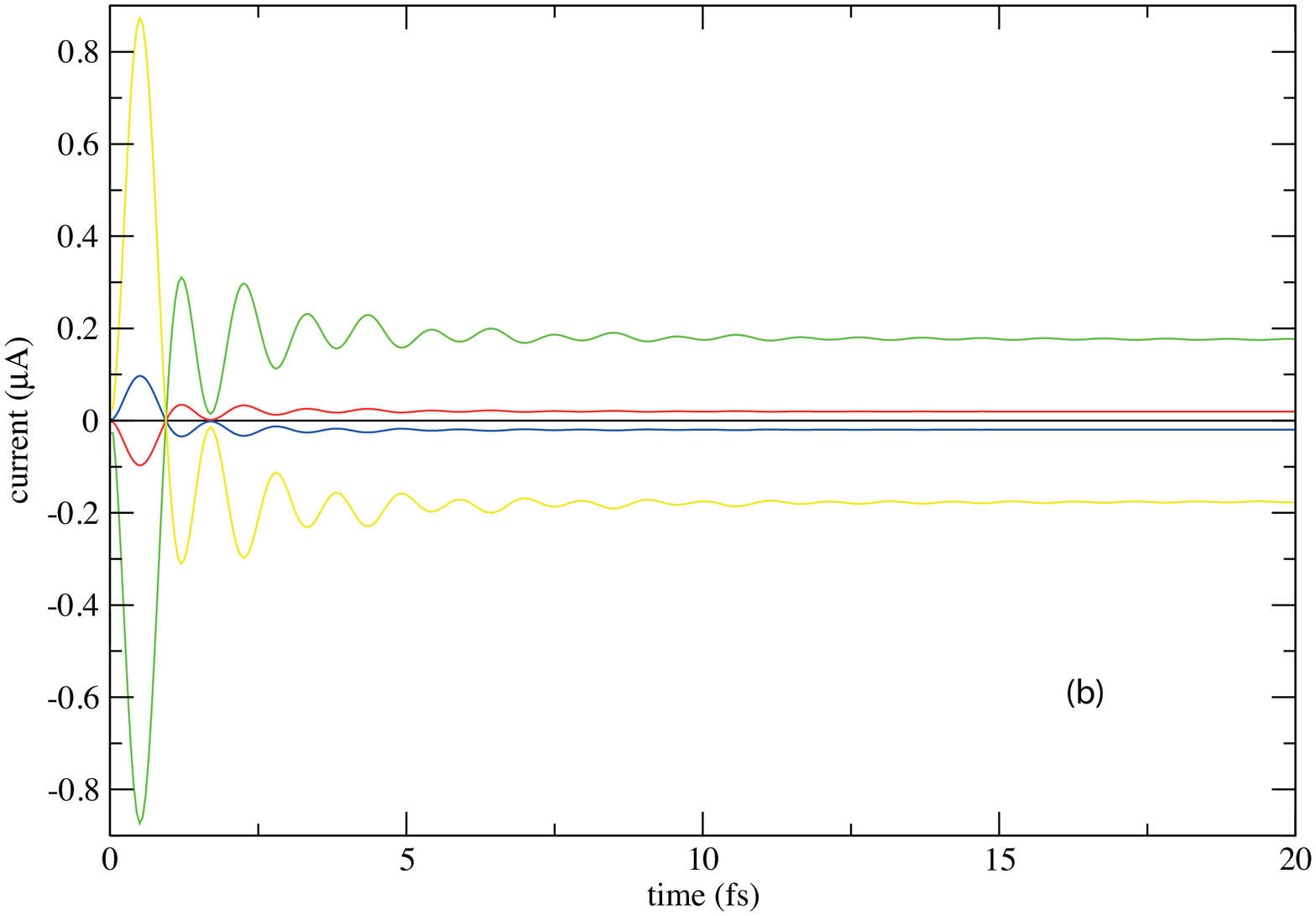}}
  \caption{(a) The left current in the nanojunction where the photon beam
    is switched on abruptly at $0~{\rm fs}$. The photon frequencies are
    $400~{\rm THz}$ (red line), $450~{\rm THz}$ (green line), $500~{\rm THz}$
    (blue line), $550~{\rm THz}$ (yellow line), and $600~{\rm THz}$ (brown
    line). The Electron-photon coupling is $M_{e\gamma} = 0.5~{\rm eV}$. In
    (b), the left (red line) and right (blue line) currents when
    $M_{e\gamma} = 0.1~{\rm eV}$ and the left (green line) and right (yellow
    line) currents when $M_{e\gamma} = 0.3~{\rm eV}$. The photon frequency is
    $500~{\rm THz}$ and $V_b = 0$.}
  \label{fig:electronphoton}
\end{figure}

In the nanojunction with an incident photon beam, the photon parameters we
use are $\chi^{\rm L} = \chi^{\rm R} = \chi^{\rm C} = 0$ and $u^{\rm L} =
u^{\rm R} = u^{\rm C} = u^{\rm LC} = u^{\rm RC} = 0.2~{\rm eV}$. The
electron-photon coupling strength values we investigated are $M_{e\gamma} =
0.1~{\rm eV}$, $0.3~{\rm eV}$, and $0.5~{\rm eV}$. Shown in
Fig.~\ref{fig:electronphoton}(a) are the left current values when the
frequencies of the incident photons are varied while the electron-photon
coupling is at $M_{e\gamma} = 0.5~{\rm eV}$. There is no source-drain bias
potential. The electron-photon is switched on in the form of a
Heaviside step function at time $t = 0$. Notice that when the interaction is
switched on the left current dips to a negative value, rises up, oscillates,
and then eventually settles down to a steady value. Also notice that the
value and direction of flow of the steady-state current depends on the
frequency of the incident photons. Shown in Fig.~\ref{fig:electronphoton}(b)
are the left and right currents when $M_{e\gamma} = 0.1~{\rm eV}$ and
$0.3~{\rm eV}$. Notice that the left current is exactly the opposite of the
right current. This means that the left and right currents would both flow
into the channel at the same time or both flow out of the channel at the
same time, maintaining a zero net current within the channel.

%--------------------------------------------------------------------------
%                        Summary and Conclusion
%--------------------------------------------------------------------------
\section{SUMMARY AND CONCLUSION}
\label{sec:summary}

We are able to calculate the time-dependent current for nanojunctions with
dynamical electron-phonon and electron-photon interactions. We use the
two-time nonequilibrium Green's functions to calculate the current. We find
that the current takes time to react to a sudden change in the device. The
current would overshoot, oscillate, and then settle down to a steady value.
Electron-phonon interactions drive the electrons out of the channel and
result in a net resistance to the current flow whenever a source-drain bias
potential is present. Electron-photon interactions also pushes the electron
current to flow. The direction of flow, however, depends on the frequency
of the incident photons. Furthermore, the direction of current flow in one
lead is opposite to the direction of current flow in the other lead, thereby
resulting in no net current within the channel.

%--------------------------------------------------------------------------
%                              Acknowledgments
%--------------------------------------------------------------------------
\section{ACKNOWLEDGMENTS}
\label{sec:acknowledgments}

CML would like to acknowledge support from the Accelerated Science and
Technology Human Resources Development Program of DOST-SEI.

%--------------------------------------------------------------------------
%                                 References
%--------------------------------------------------------------------------

\end{document}